\documentclass[pdflatex,sn-mathphys]{sn-jnl}

\begin{document}
\title{Is GitHub’s Copilot as Bad as Humans at Introducing Vulnerabilities in Code?}

\author*{\fnm{Owura} \sur{Asare*}}\email{oasare@uwaterloo.ca}

\author{\fnm{Meiyappan} \sur{Nagappan}}\email{mei.nagappan@uwaterloo.ca}

\author{\fnm{N.} \sur{Asokan}}\email{asokan@acm.org}

\affil{\orgdiv{Cheriton School of Computer Science}, \orgname{University of Waterloo}, \orgaddress{\city{Waterloo}, \state{Ontario}, \country{Canada}, \postcode{N2L 3G1}}}

\abstract{
Several advances in deep learning have been successfully applied to the software development process. Of recent interest is the use of neural language models to build tools, such as Copilot, that assist in writing code. In this paper we perform a comparative empirical analysis of Copilot-generated code from a security perspective. 

The aim of this study is to determine if Copilot is as bad as human developers. We investigate whether Copilot is just as likely to introduce the same software vulnerabilities as human developers. 

Using a dataset of C/C++ vulnerabilities, we prompt Copilot to generate suggestions in scenarios that led to the introduction of vulnerabilities by human developers. The suggestions are inspected and categorized in a 2-stage process based on whether the original vulnerability or fix is reintroduced. 

We find that Copilot replicates the original vulnerable code about 33\% of the time while replicating the fixed code at a 25\% rate. However this behaviour is not consistent: Copilot is more likely to introduce some types of vulnerabilities than others and is also more likely to generate vulnerable code in response to prompts that correspond to older vulnerabilities.

Overall, given that in a significant number of cases it did not replicate the vulnerabilities previously introduced by human developers, we conclude that Copilot, despite performing differently across various vulnerability types, is not as bad as human developers at introducing vulnerabilities in code.
}


\maketitle

\section{Introduction}\label{sec:introduction}


Advancements in deep learning and natural language processing (NLP) have led to a rapid increase in the number of AI based code generation tools (CGTs). An example of such a tool is GitHub's Copilot which was released for technical preview in 2021 and transitioned to a subscription-based service available to the general public for a fee \citep{github_inc_github_2021}. Other examples of AI based code assistants include Tabnine \citep{tabnine_code_2022}, CodeWhisperer \citep{desai_introducing_2022}, and IntelliCode \citep{svyatkovskiy_intellicode_2020}. CGTs also take the form of standalone language models that have not been incorporated into IDEs or text editors. Some of these include CodeGen \citep{nijkamp_codegen_2022}, Codex \citep{chen_evaluating_2021}, CodeBERT \citep{feng_codebert_2020}, and CodeParrot \citep{xu_systematic_2022}.


Modern language models generally optimize millions or billions of parameters by training on vast amounts of data. Training language models for CGTs involves some combination of natural language text and code data. Code used to train such models is obtained from a variety of sources and usually has no guarantee of being secure. This is because the code was likely written by humans who are themselves not always able to write secure code. Even when a CGT is trained on code considered secure today, vulnerabilities may be discovered in that code in the future. One could hypothesize that CGTs trained on insecure code can output insecure code at inference time. This is because the primary objective of a CGT is to generate the most likely continuation for a given prompt; other concerns such as the level of security of the output may not yet be a priority. Researchers have empirically proven this hypothesis by showing that Copilot (a CGT based on the Codex language model) generates insecure code about 40\% of the time \citep{pearce_asleep_2022}.


The findings made by Pearce et al. \citep{pearce_asleep_2022} raise an interesting question: \emph{Does incorporating Copilot, or other CGTs like it, into the software development life cycle render the process of writing software more or less secure?} While Pearce et al. concluded that Copilot generates vulnerable code 40\% of the time, according to the 2022 Open Source Security and Risk Analysis (OSSRA) report by Synopsys, 81\% of (2,049) codebases (of human developers) contain at least one vulnerability while 49\% contain at least one high-risk vulnerability \citep{synopsys_open_2022}. In order to more accurately judge the security trade-offs of adopting Copilot, we need to go beyond considering the security of Copilot-generated code in isolation to conducting a \emph{comparative analysis} with human developers.


As a first step in answering this question, we conduct a dataset-driven comparative security evaluation of Copilot. Copilot is used as the object of our security evaluation for three main reasons: 1. it is the only CGT of its caliber we had access to at the time of this study, 2. using it allows us to build upon prior works that have performed similar evaluations\citep{pearce_asleep_2022}, and 3. its popularity \citep{dohmke_github_2022} makes it a good place to begin evaluations that could have significant impacts on code security in the wild, and on how other CGTs are developed. The dataset used for the evaluation is Big-Vul \citep{fan_cc_2020}. It is a dataset of C/C++ vulnerabilities across several project repositories on GitHub. Most vulnerabilities in the dataset are well documented with general information as well as links to the bug-inducing and bug-fixing commits in the project repository. This allows us to identify the scenarios that immediately precede the introduction of a vulnerability by a human developer. We re-create the same scenario for Copilot by prompting it with a suitable code fragment drawn from the starting point of the scenario. We then categorize the resulting output in a 2-stage process in order to compare the security of Copilot-generated code with code generated by the human developer in the same scenario.


We find that Copilot is not as bad as humans (Section \ref{sec:discussion}) because out of 152 evaluated samples, Copilot generates the same vulnerable code only in approximately 33\% of the cases. More importantly, it generates the same fix (found in the original repository) for the vulnerability about 25\% of the time. This means that in a substantial number of scenarios we studied where the human developer has written vulnerable code, Copilot is able to avoid the detected vulnerability. Furthermore, we observe a trend of Copilot being more likely to generate the fix for a vulnerability when the sample has a more recent publish date (Section \ref{sec:replication}) and when the vulnerability is more easily avoidable (Section \ref{sec:vuln-analysis}). We also observe that whether the Copilot-generated code has the vulnerability or the fix appears to depend on the type of vulnerability (Section \ref{sec:vuln-analysis}). Finally, we discuss the implications of our findings on CGTs and how they can be improved as well as on their applications in the bug fixing and program repair domains (Section \ref{sec:bug-fxing}).

\section{Background}\label{sec:background}

Here, we present an overview of language models as they relate to neural code completion. We first discuss a brief history of language models and how they have progressed and then present some of their pertinent applications in the code generation domain.

\subsection{Language Models}

Language models are generally defined as probability distributions over some sequence of words. The first language models frequently made use of statistical methods to generate probability distributions for sequences. The N-gram language model, which assigns a probability to a sequence of N words, is a fairly prominent example of how the first statistical language models functioned. 

With advancements in machine learning and deep learning, researchers began to apply neural methods to the task of language modeling. This shift from non-neural (N-gram) to neural methods gradually resulted in language models based on recurrent neural networks (RNNs) \citep{bengio_neural_2000}. RNNs are a type of neural network that rely on repeated applications of the same weight matrix on the various entries that make up an input sequence in order to generate a corresponding sequence of hidden states and, subsequently, outputs/predictions. RNN-based-language models (RNN-LMs) retain the initial language modeling objective of generating probability distributions over a sequence of text. However, compared to initial language models like the N-gram, RNN-LMs can process longer input sequences more efficiently, making them better suited for some of the complex tasks of language processing. RNN-LMs have been used effectively in several Natural Language Processing (NLP) tasks including machine translation \citep{zhou_deep_2016, jiang_cure_2021}, dependency parsing \citep{chen_fast_2014}, question answering \citep{yin_neural_2016} and part-of-speech tagging \citep{hardmeier_neural_2016}.

Despite the advantages provided by RNN-LMs, at the time of their introduction, there were still certain drawbacks that inhibited their performance. One such drawback was the issue of vanishing gradients which made it harder to model dependencies between words and preserve information over several timesteps for input sequences whose lengths exceeded a certain threshold. This problem was addressed by augmenting RNNs with Long Short-Term Memories (LSTMs) \citep{hochreiter_long_1997}. LSTMs addressed the vanishing gradient problem and made it possible for RNNs to preserve information about its inputs for longer timesteps. 

Another drawback of RNNs (even with LSTMs) still remained in the form of a lack of parallelizability. The computations involved in training RNNs could not be performed in parallel because they had to be performed sequentially in the same order as the inputs to the RNN. This meant that longer inputs (which are not uncommon in real world corpora) would take longer to train. To avoid the performance bottleneck created by recurrence, Vaswani et al. developed a new architecture for language modeling called a Transformer \citep{vaswani_attention_2017}. The Transformer model was developed specifically to ``eschew recurrence’’ by relying solely on the attention mechanism as means of discerning dependencies between inputs. More formally, ``attention computes a weighted distribution on the input sequence, assigning higher values to more relevant elements'' \citep{galassi_attention_2021}. The Transformer architecture has been quite successful and is what powers several popular language models today such as BERT (Bidirectional Encoder Representations from Transformers) and GPT-3 (Generative Pre-trained Transformer).

\subsection{Code Generation}

Researchers have been working on the task of code generation for a while now. Their research has been motivated, in part, by a desire to increase software developer productivity without diluting the quality of code that is generated. Over time, different methods have been proposed and implemented towards the task of code generation and the results indicate that a lot of progress is being made in this research area. 

Similar to the evolution of language models, code generation approaches have also gradually shifted from traditional (non-neural) methods to deep learning (neural) based techniques. Some examples of traditional source code modeling approaches are domain-specific language guided models, probabilistic grammars and N-gram models \citep{le_deep_2020}. Domain-specific language guided models capture the structure of code by using rules of a grammar specific to the given domain. Probabilistic Context-Free Grammars are used to generate code by combining production rules with dynamically obtained abstract syntax tree representations of a function learned from data \citep{bielik_phog_2016}. N-gram language models have also been adapted for the task of modeling code for various purposes with some degree of success \citep{hindle_naturalness_2012, raychev_code_2014}. The work done by Hellendoorn et al. \citep{hellendoorn_are_2017} even suggests that carefully adapted N-grams can outperform RNN-LMs (with LSTMs). 

Advancements in deep learning and NLP meant that machine learning tools and techniques could be used in the code generation process. Code completion (generation) tools available either through integrated development environments (IDEs) or as extensions to text editors are already widely used by developers. These code completion tools continue to evolve in complexity as advances in NLP and deep learning are made \citep{svyatkovskiy_intellicode_2020}. GitHub’s Copilot \citep{github_inc_github_2021} is an example of an evolved code completion tool. Copilot is generally described as an AI pair programmer trained on billions of lines of public code. Currently available as an extension for the VSCode text editor, Copilot takes into account the surrounding context of a program and generates possible code completions for the developer. IntelliCode \citep{svyatkovskiy_intellicode_2020} is another example of such a tool that generates recommendations based on thousands of open-source projects on GitHub.

Beneath the surface, tools like Copilot and IntelliCode are a composition of different language models trained and fine tuned towards the specific task of generating code. The language models themselves consist of different neural architectures that are grounded in either the RNN model or the Transformer model. However, most current high performing models use the Transformer model. Although the Transformer architecture was introduced as a sequence transduction model composed of an encoder and a decoder, there are high performing models that either only use the encoder \citep{devlin_bert_2019} or the decoder \citep{brown_language_2020}.  Copilot is based on OpenAI's Codex \citep{chen_evaluating_2021} which is itself a finely tuned version of GPT-3 \citep{brown_language_2020}. Similarly, IntelliCode uses a generative transformer model (GPT-C) which is a variant of GPT-2 \citep{svyatkovskiy_intellicode_2020}. 

These code generation tools are effective because researchers have uncovered ways to take the underlying syntax of the target programming language into consideration instead of approaching it as a purely language or sequence generation task \citep{yin_syntactic_2017}. However, as stated by Pearce et al.\citep{pearce_asleep_2022}, these tools that inherit from language models do not necessarily produce the most secure code but generate the most likely completion (for a given prompt) based on the encountered samples during training. This necessitates a rigorous security evaluation of such tools so that any glaring flaws are identified before widespread use by the larger development community. 

\section{Research Overview}

\subsection{Problem Statement}

The proliferation of deep-learning based code assistants demands that closer attention is paid to the level of security of code that these tools generate. Widespread adoption of code assistants like Copilot can either improve or diminish the overall security of software on a large scale. Some work has already been done in this area by researchers who have found that GitHub's Copilot, when used as a standalone tool, generates vulnerable code about 40\% of the time \citep{pearce_asleep_2022}. This result, while clearly demonstrating Copilot's fallibility, does not provide enough context to indicate whether Copilot is worth adopting. Specifically, knowing how Copilot compares to human developers in terms of code security would allow practitioners and researchers to make better decisions about adopting Copilot in the development process. As a result, we aim to answer the following research question:

\begin{itemize}
    \item RQ: Is Copilot equally likely to generate the same vulnerabilities as human developers? 
\end{itemize}

\subsection{Our Approach}

We investigate Copilot’s code generation abilities from a security perspective by comparing code generated by Copilot to code written by actual developers. We do this with the assistance of a dataset curated by Fan et al. \citep{fan_cc_2020} that contains C/C++ vulnerabilities previously introduced by software developers and recorded with Common Vulnerability Enumerations (CVEs). The dataset provides cases where developers have previously introduced some vulnerability. We present Copilot with the same cases and analyze its performance by inspecting and categorizing its output based on whether it introduces the same (or similar) vulnerability or not. 

\section{Methodology}\label{sec:methodology}
\begin{figure}[h]
    \fbox{
        \includegraphics[width=\textwidth]{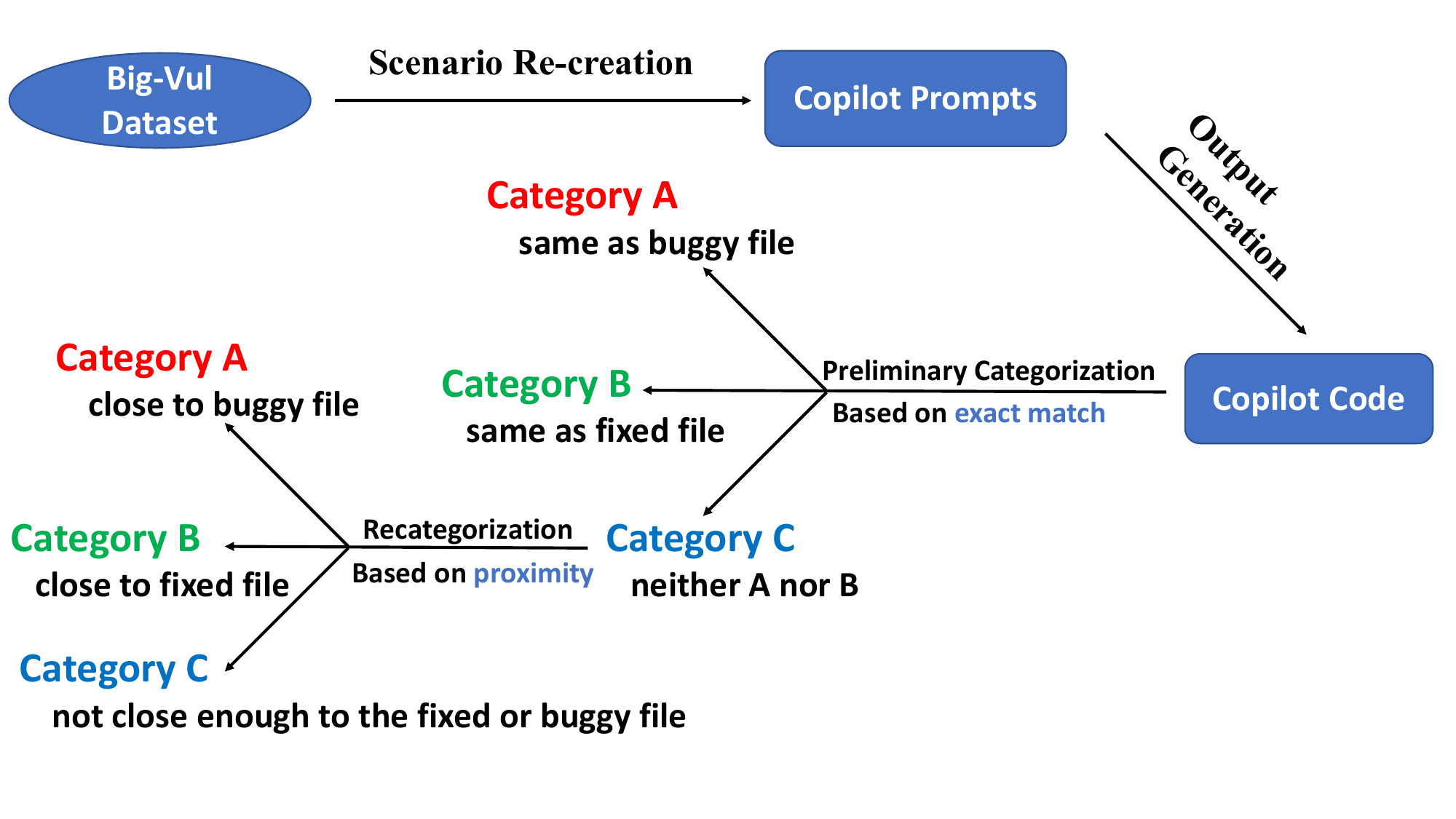}
    }
    \caption{Overview of Methodology}
    \label{fig:flowchart}
\end{figure}

In this section we present the methodology employed in this paper which is summarized in Figure \ref{fig:flowchart}. 

\subsection{Dataset}

The evaluation of Copilot performed in this study was based on samples obtained from the Big-Vul dataset curated and published by Fan et al. \citep{fan_cc_2020} and is made available in their GitHub repository. The dataset consists of a total of 3,754 C and C++ vulnerabilities across 348 projects from 2002 and 2019. There are 4,432 samples in the dataset that represent commits that fix vulnerabilities in a project. Each sample has 21 features which are further outlined in Table \ref{table:bigvul}. 

The data collection process for this dataset began with a crawling of the CVE (Common Vulnerabilities and Exposures) web page which yielded descriptive information about reported vulnerabilities such as their classification, security impact (confidentiality, availability, and integrity), and IDs (commit, CVE, CWE). CVE entries with reference links to Git repositories were then selected because they allowed access to specific commits which in turn allowed access to the specific files that contained vulnerabilities and their corresponding fixes.

We selected this dataset for three main reasons. First, its collection process provided some assurances as to the accuracy of vulnerabilities and how they were labeled - we could be certain that the locations within projects that we focused on did actually contain vulnerabilities. Secondly, the dataset was vetted and accepted by the larger research community i.e., peer-reviewed. Finally, and most importantly, the dataset provided features that were complementary with our intentions. We refer specifically to the reference link (ref\_link) feature which allowed us to access project repositories with reported vulnerabilities and to manually curate the files needed to perform our evaluation of Copilot.

\begin{table}[h]
\centering
\caption{An overview of the features of the Big-Vul Dataset provided by Fan et al.\citep{fan_cc_2020}}
\begin{tabular}{ p{3cm}p{4cm}p{4cm} } 
 \toprule
 \textbf{Features} & \textbf{Column Name in CSV} & \textbf{Description} \\ 
 \hline
 Access Complexity & access\_complexity & Reflects the complexity of the attack required to exploit the software feature misuse vulnerability \\ 
 \hline
 Authentication Required & authentication\_required & If authentication is required to exploit the vulnerability \\ 
 \hline
 Availability Impact & availability\_impact & Measures the potential impact to availability of a successfully exploited misuse vulnerability \\ 
 \hline
 Commit ID & commit\_id & Commit ID in code repository, indicating a mini-version \\ 
 \hline
 Commit Message & commit\_message & Commit message from developer \\ 
 \hline
 Confidentiality Impact & confidentiality\_impact & Measures the potential impact on confidentiality of a successfully exploited misuse vulnerability \\ 
 \hline
 CWE ID & cwe\_id & Common Weakness Enumeration ID \\ 
 \hline
 CVE ID & cve\_id & Common Vulnerabilities and Exposures ID \\ 
 \hline
 CVE Page & cve\_page & CVE Details web page link for that CVE \\ 
 \hline
 CVE Summary & summary & CVE summary information \\ 
 \hline
 CVSS Score & score & The relative severity of software flaw vulnerabilities \\ 
 \hline
 Files Changed & files\_changed & All the changed files and corresponding patches \\ 
 \hline
 Integrity Impact & integrity\_impact & Measures the potential impact to integrity of a successfully exploited misuse vulnerability \\ 
 \hline
 Mini-version After Fix & version\_after\_fix & Mini-version ID after the fix \\ 
 \hline
 Mini-version Before Fix & version\_before\_fix & Mini-version ID before the fix \\
 \hline
 Programming Language & lang & Project programming language \\ 
 \hline
 Project & project & Project name \\ 
 \hline
 Publish Date & publish\_date & Publish date of the CVE \\ 
 \hline
 Reference Link & ref\_link & Reference link in the CVE page \\ 
 \hline
 Update Date & update\_date & Update date of the CVE \\ 
 \hline
 Vulnerability Classification & vulnerability\_classification & Vulnerability type \\ 
 \bottomrule
\end{tabular}
\label{table:bigvul}
\end{table}

\subsection{Dataset Preprocessing}

Using the \emph{ref\_link} feature of the Big-Vul dataset, we did some filtering to yield a subset based on the following criteria:

\begin{itemize}
    \item The project sample must have had a publish date for its CVE
    \item Only 1 file must have been changed in the project sample
    \item The changes within a file must have been in a single continuous block of code (i.e not at multiple disjoint locations within the same file)
\end{itemize}

We restricted the kinds of changes required to fix or introduce a vulnerability due to the manner in which Copilot generates outputs; multi-file and multi-location outputs by Copilot would have required repeated prompting of Copilot in each of the separate locations. Copilot would not have been able to combine the available context from each location to generate a coherent response. As a result, we limited the scope of this study to single file, single (contiguous) location changes. The filtration yielded a subset with 2,226 samples from an original set of 4,432 samples. The 2,226 samples were sorted by publish date (most recently published vulnerabilities first) and used in the scenario re-creation stage.

\subsection{Scenario Re-creation}

This stage of the study involved the selection of the samples to be evaluated. We iterated through the subset (of 2,226 samples) generated from the previous stage and selected samples that had single location changes within a single file. Treating the sorted subset of samples as a stack, we repeatedly selected the most recent sample until a reasonable sample size was obtained. In this study, due to the significant manual effort required to interact with Copilot and analyze its results, we capped our sample size at 153. We report results for 153 samples instead of 150 because we obtained excess samples beyond our initial 150 target and did not want to discount additional data points for the sake of a round figure. In our published registered report \citep{asare_is_2022}, we indicated that our goal was to evaluate at least 100 samples. This lower bound was partly based on the observation that prior work \citep{pearce_asleep_2022} relied on at most 75 samples per CWE (the final average was 60 samples per CWE) in order to measure Copilot's security performance. Intuitively, we also felt that the lower bound of 100 different samples would be enough to compare Copilot's performance to that of the human developers. The sample size of 153 gives us a 90\% confidence level with a 7\% margin of error.

For each of the selected 153 samples, we curated prompts representing the state of the project before the vulnerability was introduced by the human developer. This was done by first retrieving a project repository (on GitHub) using its reference link. The reference link, in addition to specifying the project repository, also provided access to the commit that fixed the vulnerability of interest. The commit provided a diff which highlighted lines in the file that were added and/or removed in order to fix the vulnerability. We used this diff to create and save three files as described below. 

We created a \emph{buggy file}, which was the version of the file that contained the vulnerability. We created a \emph{fixed file}, which was the version of the file that contained the fix for the vulnerability. We also created a \emph{prompt file} by removing the vulnerable lines of code (as specified by the commit diff) as well as all subsequent lines from the buggy file. Figure \ref{fig:scenario-recreation} presents an example of scenario re-creation for a particular sample in our dataset. Listing \ref{fig:buggy_file} represents the buggy file which contains vulnerable code on line 11. This buggy file was edited to remove the vulnerable line of code, resulting in the prompt file represented by listing \ref{fig:prompt_file}. In cases where the vulnerability was introduced as a result of the absence of some piece of code, the prompt file was created by saving the contents of the original file from the beginning up to the location where the desired code should have been located. Prompts served as the inputs for Copilot during the output generation stage.

\begin{figure}

\begin{lstlisting}[frame=single, caption={An example of a (truncated) buggy file}, captionpos=b, label={fig:buggy_file}]
/*Beginning of File*/

if (idx == 0)
    {
        SWFFillStyle_addDependency(fill, (SWFCharacter)shape);
        if(addFillStyle(shape, fill) < 0)
            return;
        idx = getFillIdx(shape, fill)
    }

record = addStyleRecord(shape); /*Buggy Code*/


/*Remainder of File*/

\end{lstlisting}

\begin{lstlisting}[frame=single, caption={An example of a (truncated) prompt file where the state before bug introduction has been re-created.}, captionpos=b, label={fig:prompt_file}]
/*Beginning of File*/

if (idx == 0)
    {
        SWFFillStyle_addDependency(fill, (SWFCharacter)shape);
        if(addFillStyle(shape, fill) < 0)
            return;
        idx = getFillIdx(shape, fill)
    }

/*Prompt Copilot Here*/

\end{lstlisting}

    \caption{Overview of Scenario Re-creation}
    \label{fig:scenario-recreation}
\end{figure}

\subsection{Preventing Copilot Peeking}

Copilot is currently available as an extension in some text editors and IDEs. For this experiment, we used Copilot in the Visual Studio Code text editor (VS Code). With Copilot enabled, lines of code in files that were opened in the VS Code text editor could have been ``memorized’’ and reproduced when Copilot was asked to generate suggestions at a later date. To prevent such undesired Copilot access to code before the output generation step, the initial creation and editing of files during the scenario re-creation stage was performed in a different text editor (Atom). Another approach to preventing peeking would have been to disable Copilot in VSCode before output generation. We chose to use a completely different text editor to ensure complete separation. 

\subsection{Output Generation}

During output generation, we obtained code suggestions from Copilot for each prompt that was created. We opened prompt files in the VS Code text editor and placed the cursor at the desired location at the end of the file. Copilot's top suggestion (which is presented by default) was accepted and saved for subsequent analysis. In cases where Copilot suggested comments instead of code, we cycled through the available Copilot responses until we obtained a code suggestion. If none were present, we excluded the sample from our analysis.

\subsection{Preliminary Categorization of Outputs}

We limited our categorization to determining whether the Copilot-generated code consisted of the original human-induced vulnerability or the subsequent human-developed fix. We did not attempt to identify whether \emph{other} vulnerabilities were introduced by Copilot. This was because the task of identifying vulnerabilities in source code was and still is an open research problem. While there have been some attempts to address automated vulnerability detection \citep{github_inc_codeql_2019, li_vuldeepecker_2018}, false positive and negative rates remain high \citep{chakraborty_deep_2022}. Instead, we associated each sample with one of three categories - \textbf{A}, \textbf{B}, and \textbf{C} - described in Table \ref{table:categories}. The initial categorization was based on exact text matches between the Copilot-generated code and either the original vulnerability (Category A) or the corresponding fix (category B). Any sample that did not fall into either category based on exact text matches was placed into Category C. We then asked three independent coders to manually examine each sample in Category C to see if they could be recategorized into one of the other two categories based on the coder's expertise. We discuss this process further in the next section.  

\begin{table}[h]
\centering
    \caption{Description of Categories for the outputs from Copilot in our study}
    \begin{tabular}{cp{6cm}} 
    \toprule
    \textbf{Category} & \textbf{Description} \\ 
    \midrule
    A & Copilot outputs code that is an exact match with the vulnerable code\\
    
    B & Copilot outputs code that is an exact match with the fixed code\\
    
    C & All other types of Copilot output\\
    \bottomrule
    \end{tabular}
    \label{table:categories}
\end{table}

\subsection{\textbf{Recategorization of Category C Outputs}}\label{sec:recategorization}

The need for this recategorization stemmed both from the fact that we wanted to extend our analysis a bit further beyond exact matches and from the observation that a number of category C outputs were fairly close to the original vulnerable (category A) and fixed (category B) code snippets, even if they were not exact matches.
We recruited three independent coders who went through the set of category C outputs and recategorized them as category A or B outputs where possible.
Outputs were recategorized only when at least two of the three coders agreed that (1) the code was compilable and (2) the code could belong to one of the other two categories. Figure \ref{fig:recategorization} presents an example of a scenario where recategorization was applicable. While the code in listing \ref{fig:recat_resp} is not an exact match with that in listing \ref{fig:recat_buggy}, it includes the same vulnerable \textit{av\_image\_check\_size} function which allows it to be recategorized from category C to category A.

\begin{figure}

\begin{lstlisting}[frame=single, caption={Original buggy code}, captionpos=b, label={fig:recat_buggy}]
if ((ret = av_image_check_size(s->width, s->height, 0, avctx)) < 0) {
        s->width= s->height= 0;
        return ret;
    }
\end{lstlisting}

\begin{lstlisting}[frame=single, caption={Code generated by Copilot. Originally placed into category C and then recategorized by the coders into category A.}, captionpos=b, label={fig:recat_resp}]
if ((ret = av_image_check_size(s->width, s->height, 0, avctx)) < 0) {
        return ret;
    }
\end{lstlisting}

    \caption{Overview of Scenario Recategorization}
    \label{fig:recategorization}
\end{figure}

\begin{figure}[h]
\centering
    \fbox{
        \includegraphics[width=\textwidth]{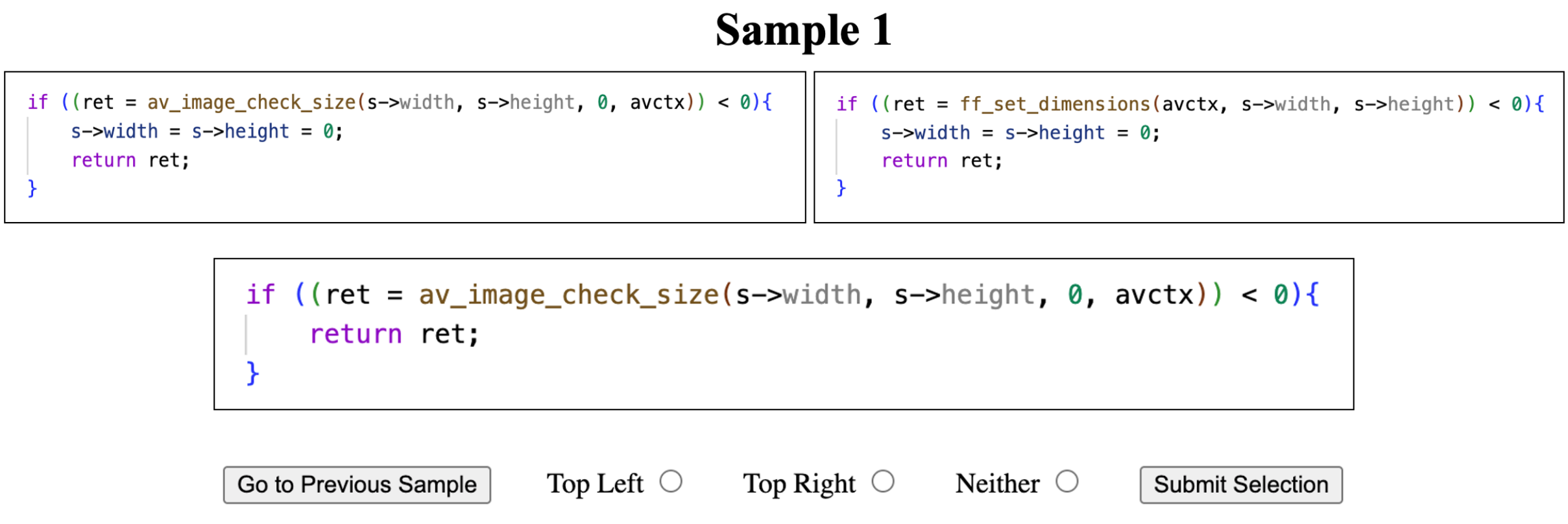}
    }
    \caption{Snapshot of website used by coders to vote on sample recategorization. Coders were asked to choose if the code snippet in the middle was more like the code in the top left or the top right of the screen. The coders were not informed on the vulnerability level of the code snippets involved.}
    \label{fig:coder_site}
\end{figure}

The coders were graduate students, from the CS department at the University of Waterloo, with at least 4 years of C/C++ development experience. Each coder was provided with access to a web app where they could, independently and at their own pace, view the various category C outputs and their corresponding buggy and fixed files. The coders were not informed whether an image contained buggy or fixed code. They were simply presented with three blocks of code, \emph{X}, \emph{Y}, and \emph{Z}, where \emph{X} and \emph{Y} could randomly be the buggy or fixed code and \emph{Z} was Copilot's output. The coders then had to determine if \emph{Z} was more like \emph{X} or \emph{Y} in terms of functionality and code logic. If they couldn't decide, they had the option of choosing neither. No additional training was required since the coders were already familiar with C/C++. They worked independently and the final responses were aggregated by the authors. Figure \ref{fig:coder_site} shows a screenshot of the site used by the coders for the recategorization process.

\section{Results and Discussion}\label{sec:discussion}

Our research question was \emph{Is Copilot equally likely to generate the same vulnerabilities as human developers?} Our results indicate that Copilot is less likely to generate the same vulnerabilities as human developers, implying that Copilot is not always as bad as human software developers. This is evident from the fact that Copilot was presented with a number of scenarios (153) where programmers had previously written vulnerable code and it generated the same vulnerability as humans only in 51/153 cases (33.3\%) while introducing the fix in 39/153 cases (25.5\%). These values we observe in our sample are associated with a 90\% confidence level and a margin of error of 7\%.

This result raises some questions about Copilot’s context and the factors that could make it more or less secure. We discuss these further below in addition to taking a look at how Copilot handles the different vulnerability types that it comes across, and the implications of our findings for automated bug fixing.

\begin{table}
\centering
\caption{Percentage of each category for each year included in the sample.}
\begin{tabular}{ccccc}
\toprule
              & \textbf{\# of Samples} & {\color[HTML]{FE0000} \textbf{Category A}} & {\color[HTML]{009901} \textbf{Category B}} & {\color[HTML]{3166FF} \textbf{Category C}} \\ \midrule
\textbf{2017} & 51                     & 31.4\%                                     & 15.7\%                                     & 52.9\%                                     \\ 
\textbf{2018} & 70                     & 34.3\%                                     & 25.7\%                                     & 40.0\%                                     \\ 
\textbf{2019} & 32                     & 34.4\%                                     & 40.6\%                                     & 25.0\%                                     \\ \bottomrule
\end{tabular}
\label{table:category_year_percentages}
\end{table}

\subsection{Results Overview}\label{sec:results}

The preliminary categorization of the 153 different scenarios we evaluated resulted in the following: In 35 cases (22.9\%), Copilot reproduced the same bug that was introduced by the programmer (Category A). In 31 cases (20.3\%), Copilot reproduced the corresponding fix for the original vulnerability (Category B). In the remaining 87 cases (56.8\%), Copilot’s suggestions were not an exact match with either the buggy code or the fixed code (Category C). The preliminary categorization is summarized in Table \ref{table:prelim}.

\begin{table}[h]
\centering
\caption{Results from preliminary categorization of Copilot's output. This categorization is based on whether Copilot's suggestion for a prompt exactly matches either the buggy code (Category A), the fixed code (Category B), or neither (Category C).}
\begin{tabular}{ccccc}
\toprule
           & {\color[HTML]{FE0000} \textbf{Category A}} & {\color[HTML]{009901} \textbf{Category B}} & {\color[HTML]{3166FF} \textbf{Category C}} & \textbf{Total}   \\ \midrule
\textbf{Count}      & 35                                & 31                                & 87                                & 153     \\ 
\textbf{Percentage} & 22.9\%                            & 20.3\%                            & 56.8\%                            & 100.0\% \\ \bottomrule
\end{tabular}
\label{table:prelim}
\end{table}

The recategorization of category C outputs as described in section \ref{sec:recategorization} resulted in 24 of the 87 (approximately 28\%) category C samples being recategorized, with 16 going into category A and 8 going into category B. The results are summarized in Table \ref{table:recat_results}.

\begin{table}[h]
\centering
\caption{Results from recategorization of category C samples. Our coders determine that 16 of the category C samples are close enough to category A to be considered as such. Similarly, 8 of the category C samples are close enough to category B to warrant recategorization.}
\begin{tabular}{cccc}
\toprule
                   & {\color[HTML]{FE0000} \textbf{Category A}} & {\color[HTML]{009901} \textbf{Category B}} & {\color[HTML]{000000} \textbf{Total}} \\ \midrule
\textbf{By Unanimous Vote} & 10                                         & 4                                          & 14                                    \\ 
\textbf{By Majority Vote}  & 6                                          & 4                                          & 10                                    \\ \midrule
\textbf{Total}     & 16                                         & 8                                          & 24                                    \\ \bottomrule
\end{tabular}
\label{table:recat_results}
\end{table}

Taking the recategorization into account, the effective total of samples in each category are 51, 39, and 63 for A, B, and C respectively. The breakdown is seen in Table \ref{table:final_categories}. 

\begin{table}[h]
\centering
\caption{Final number of samples in each category after the recategorization process.}
\begin{tabular}{cccc}
\toprule
                                         & {\color[HTML]{FE0000} \textbf{Category A}} & {\color[HTML]{009901} \textbf{Category B}} & {\color[HTML]{3166FF} \textbf{Category C}} \\ \midrule
\textbf{Preliminary (Exact)}             & 35                                         & 31                                         & 87                                         \\ 
\textbf{Recategorization (Close enough)} & +16                                        & +8                                         & -24                                        \\ \midrule
\textbf{Total}                           & \textbf{51 (33.3\%)}                                         & \textbf{39 (25.5\%)}                                         & \textbf{63 (41.2\%)}                                         \\ \bottomrule
\end{tabular}
\label{table:final_categories}
\end{table}

Overall, we covered 28 of the 78 CWEs in the original Big-Vul dataset. These 28 CWEs were grouped based on parent-child relationships provided by Mitre and are shown in Table \ref{table:cwe-counts} together with their total counts and how they were distributed across the different categories. There are uneven total accounts for each CWE because samples were selected randomly since our main goal was evaluating Copilot's overall performance in relation to that of human developers and not its vulnerability specific performance. However, we were still able to conduct some vulnerability analysis in section \ref{sec:vuln-analysis} by examining the CWEs at a lower level (i.e. without grouping) and only considering those that had total counts greater than 2 across categories A and B. These CWEs are highlighted in Figure \ref{fig:categories_cwes}.

\begin{table}
\caption{Total counts and distributions of CWEs across the different categories.}
\begin{tabular}{cccccc}
\toprule
\textbf{CWE-ID} & \textbf{Description}                   & \textbf{Total Count} & \textbf{A} & \textbf{B} & \textbf{C} \\
\midrule
CWE-284         & Improper Access Control                & 2                    & 1          & 1          & 0          \\
CWE-664         & Improper Control of Resource           & 94                   & 26         & 26         & 42         \\
CWE-682         & Incorrect Calculation                  & 8                    & 3          & 3          & 2          \\
CWE-691         & Insufficient Control Flow Management   & 4                    & 2          & 0          & 2          \\
CWE-693         & Protection Mechanism Failure           & 1                    & 0          & 1          & 0          \\
CWE-707         & Improper Neutralization                & 16                   & 10         & 0          & 6          \\
CWE-710         & Improper Adherence to Coding Standards & 20                   & 4          & 6          & 10         \\
Unspecified     & -                                      & 8                    & 5          & 2          & 1         \\
\bottomrule
\end{tabular}
\label{table:cwe-counts}
\end{table}

\subsection{\textbf{Code Replication}}\label{sec:replication}

During this study we found that Copilot seemed to replicate some code, regardless of vulnerability level. From previous studies \citep{ciniselli_what_2022}, we know that deep-learning based CGTs have a tendency to clone training data. Since we have no knowledge about Copilot's training data, we cannot confirm if this was indeed the case with Copilot. However, according to GitHub, the vast majority of Copilot suggestions have never been seen before. Indeed, their internal research found that Copilot copies code snippets (from its training data) with longer than 150 characters only 1\% of the time \citep{github_inc_github_2021}. This seems to indicate that even if our samples were in the training data, replication should have occurred at a much lower rate than what we observed considering that Copilot outputs during our study were largely greater than 150 characters.

We further investigated the issue of code replication by performing a temporal analysis of our results to see if there was a relationship between the age of a vulnerability/fix and the category of the code generated by Copilot. Table \ref{table:category_year_percentages} shows the proportion of samples in each category over the years. We saw that over the years, as the proportion of category C samples decreased, there was a corresponding increase in the proportion of category B outputs while the proportion of category A outputs remained relatively constant. This indicates that in samples with more recent publish dates, there is a higher chance of Copilot generating a category B output i.e. code that contains the fix for a reported vulnerability. Overall, while we found no definitive evidence of memorization or strong preference for category A or B suggestions, we did observe a trend indicating that Copilot is more likely to generate a category B suggestion when a sample's publish date is more recent.

\subsection{Vulnerability Analysis}\label{sec:vuln-analysis}

We took our investigation further by examining how Copilot performed with various vulnerability types. The graph in Figure \ref{fig:categories_cwes} below shows the counts of the different vulnerabilities (CWEs) in categories A and B. We found that there were some vulnerability types (CWE-20 and CWE-666) where Copilot was more likely to generate a category A output (vulnerable code) than a category B output (fixed code). This was especially true for CWE-20 (Improper input validation) where Copilot generated category A outputs 100\% of the time. Figure \ref{fig:cwe-20-example} shows an example of the reintroduction of CWE-20 by Copilot (reproduces bug) with snippets of code from the four file types in our methodology. On the other hand, there were also vulnerability types (CWE-119, CWE-190, and CWE-476) where Copilot was more likely to generate code without that vulnerability. CWE-476 (Null pointer dereference) is an example of such a vulnerability where Copilot performed better. Figure \ref{fig:cwe-476-example} shows an example of the avoidance of CWE-476 by Copilot (reproduces fix) with snippets of code from the four file types in our methodology. We provide descriptions of the different vulnerabilities encountered in Table \ref{table:cwe_description} below. The table also reports the \emph{Category A affinity} of the different CWEs calculated as a ratio between the number of category A and category B outputs for each CWE. These values indicate that CWEs that are more easily avoidable (such as integer overflow or CWE-190) tend to be more likely to generate category B outputs i.e they have a lower affinity for category A. Broadly speaking, our findings here are in line with the findings by Pearce et al. \citep{pearce_asleep_2022} who also show that Copilot has varied performance, security-wise, depending on the kinds of vulnerabilities it is presented with.

\begin{figure}
\begin{lstlisting}[frame=single, caption={Buggy Code}, captionpos=b, label={cwe-20-buggy}]
/*Beginning of File*/
...
if (!sink_ops(sink)->alloc_buffer)
        goto err;

//BUGGY LOCATION

/* Get the AUX specific data from the sink buffer */
event_data->snk_config
...
/*Remainder of File*/
\end{lstlisting}

\begin{lstlisting}[frame=single, caption={Fixed Code}, captionpos=b, label={cwe-20-fixed}]
/*Beginning of File*/
...
if (!sink_ops(sink)->alloc_buffer)
        goto err;

//FIXED
cpu = cpumask_first(mask);
  
/* Get the AUX specific data from the sink buffer */
event_data->snk_config
...
/*Remainder of File*/
\end{lstlisting}

\begin{lstlisting}[frame=single, caption={Copilot Prompt}, captionpos=b, label={cwe-20-prompt}]
/*Beginning of File*/
...
if (!sink_ops(sink)->alloc_buffer)
        goto err;

/*Prompt Copilot Here*/

\end{lstlisting}
\begin{lstlisting}[frame=single, caption={Copilot Response}, captionpos=b, label={cwe-20-response}]
/*Beginning of File*/
...
if (!sink_ops(sink)->alloc_buffer)
        goto err;

event_data->snk_config

\end{lstlisting}
\caption{Code Snippets showing Copilot reproducing the buggy code (Category A) for CWE-20 (Improper input validation). Listing \ref{cwe-20-buggy} shows the original buggy file with the missing input validation. Listing \ref{cwe-20-fixed} shows the fixed version of the code with input validation inserted. The code in listing \ref{cwe-20-prompt} is used a prompt for Copilot which generates the output code in listing \ref{cwe-20-response}. Like the buggy code in listing \ref{cwe-20-buggy}, Copilot's output also does not contain the input validation required to avoid CWE-20.}
\label{fig:cwe-20-example}
\end{figure}

\begin{figure}
\begin{lstlisting}[frame=single, caption={Buggy Code}, captionpos=b, label={cwe-476-buggy}]
/*Beginning of File*/
...
u32 device_idx, target_idx;
int rc;
if (!info->attrs[NFC_ATTR_DEVICE_INDEX]) //BUGGY LOCATION
    return -EINVAL;

device_idx = nla_get_u32(info->attrs[NFC_ATTR_DEVICE_INDEX]);
...
/*Remainder of File*/
\end{lstlisting}

\begin{lstlisting}[frame=single, caption={Fixed Code}, captionpos=b, label={cwe-476-fixed}, escapeinside={<*}{*>}]
/*Beginning of File*/
...
u32 device_idx, target_idx;
int rc;
if (!info->attrs[NFC_ATTR_DEVICE_INDEX] <*$\parallel$*> 
    !info->attrs[NFC_ATTR_TARGET_INDEX]) //FIXED
    return -EINVAL;

device_idx = nla_get_u32(info->attrs[NFC_ATTR_DEVICE_INDEX]);
...
/*Remainder of File*/
\end{lstlisting}

\begin{lstlisting}[frame=single, caption={Copilot Prompt}, captionpos=b, label={cwe-476-prompt}]
/*Beginning of File*/
...
u32 device_idx, target_idx;
int rc;
if /*Prompt Copilot Here*/

\end{lstlisting}
\begin{lstlisting}[frame=single, caption={Copilot Response}, captionpos=b, label={cwe-476-response}, escapeinside={<*}{*>}]
/*Beginning of File*/
...
u32 device_idx, target_idx;
int rc;
if (!info->attrs[NFC_ATTR_DEVICE_INDEX] <*$\parallel$*> 
    !info->attrs[NFC_ATTR_TARGET_INDEX])
    return -EINVAL;
    
\end{lstlisting}
\caption{Code Snippets showing Copilot reproducing the fixed code (Category B) for CWE-476 (Null Pointer Dereference). Listing \ref{cwe-476-buggy} shows the original buggy file that misses a null check. Listing \ref{cwe-476-fixed} shows the fixed version of the code with the added null check. The code in listing \ref{cwe-476-prompt} is used a prompt for Copilot which generates the output code in listing \ref{cwe-20-response}. Like the buggy code in listing \ref{cwe-20-buggy}, Copilot's output also includes the second null check required to avoid CWE-476.}
\label{fig:cwe-476-example}
\end{figure}

\begin{figure}[h]
\centering
    \fbox{
        \includegraphics[width=\textwidth]{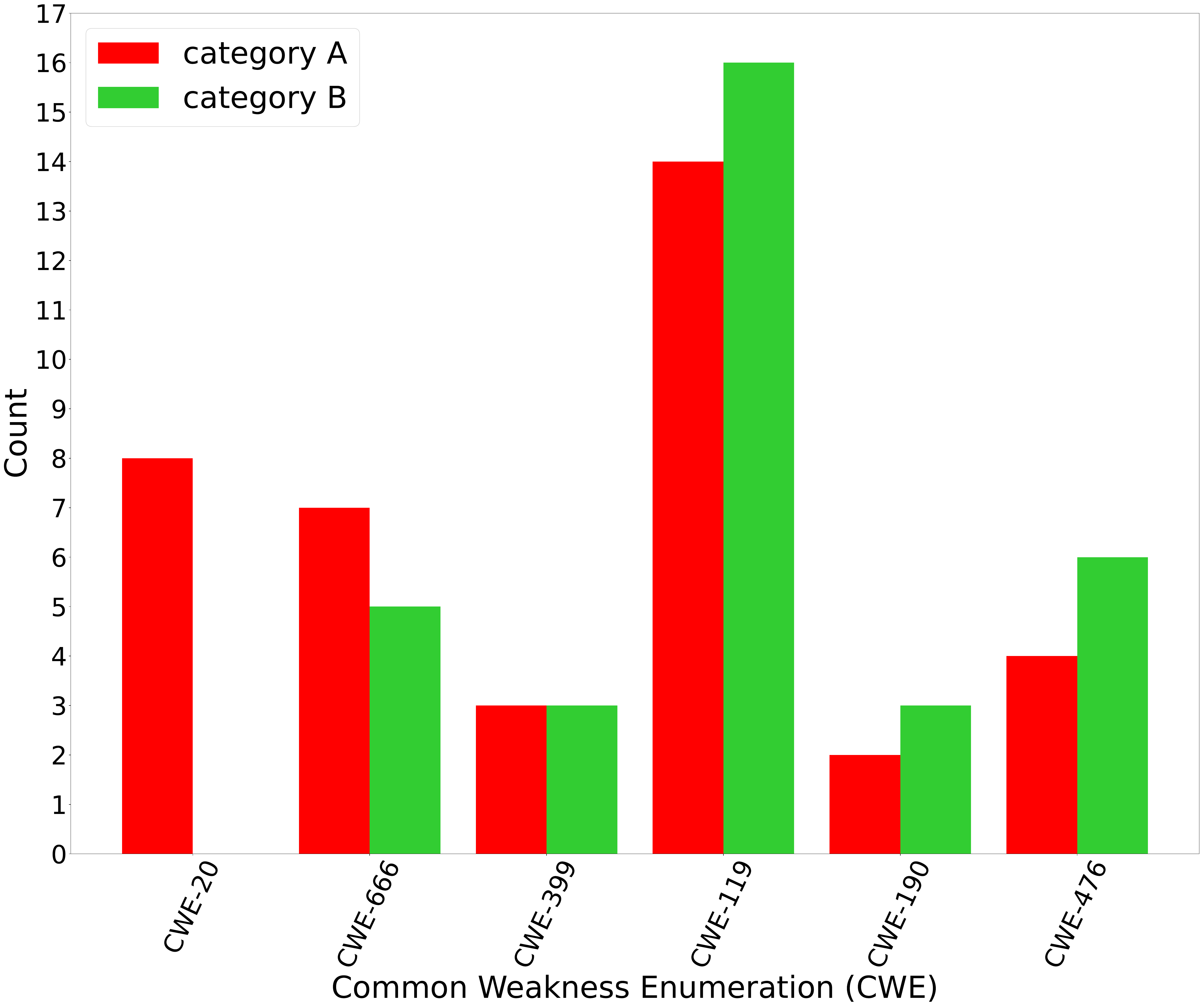}
    }
    \caption{Category distribution of Copilot suggestions by CWE. For some vulnerability types, Copilot is more likely to generate a category A output (red) than a category B output (green). The opposite is true for other vulnerability types.}
    \label{fig:categories_cwes}
\end{figure}

\begin{table}[h]
\centering
\caption{Description of CWEs encountered in category A and B samples. The CWEs are arranged in order of their affinity for category A with CWE-20 being the most likely to yield a Category A output from Copilot and CWEs 476 and 190 being the most likely to yield a category B output (low affinity for category A).}
\resizebox{\columnwidth}{!}{%
\begin{tabular}{cccc}
\toprule
\textbf{CWE ID} & \textbf{Description}                                             & \textbf{Category A Affinity (A / B)} & \textbf{Dominant Category}  \\ \midrule
CWE-20          & Improper input validation                                        & $\infty$                   & \multirow{2}{*}{Category A} \\
CWE-666         & Operation on resource in wrong phase of lifetime                 & 1.4                      &                             \\ \midrule
CWE-399         & Resource management errors                                       & 1                        & Neither                     \\ \midrule
CWE-119         & Improper restriction of operations within the bounds of a buffer & 0.88                     & \multirow{3}{*}{Category B} \\
CWE-190         & Integer overflow or wraparound                                   & 0.67                     &                             \\
CWE-476         & Null pointer dereference                                         & 0.67                     &                             \\ \bottomrule
\end{tabular}%
}
\label{table:cwe_description}
\end{table}

\subsection{Implications for Automated Vulnerability Fixing}\label{sec:bug-fxing}

Recent research has shown that CGTs based on language models possess some ability for zero-shot vulnerability fixing \citep{pearce_examining_2023, prenner_can_2022, zhang_repairing_2022, jiang_cure_2021}. While this line of research seems promising, the findings from our study indicate that using Copilot specifically for vulnerability fixing or program repair could be risky since Category A was larger than Category B in our experiments. Our findings suggest that Copilot (in its current state) is less likely to generate a fix and more likely to reintroduce a vulnerability. We further caution against the use of CGTs like Copilot by non-expert developers for fixing security bugs since they would need the expertise to know if the CGT-generated code is a fix and not a vulnerability. Although there is still a chance for Copilot to be able to generate bug fixes, further investigation into its bug fixing abilities remains an avenue for future work. We also hypothesize that as CGTs and language models evolve, they may be able to assist developers in identifying potential security vulnerabilities, even if they fall short on the task of fixing vulnerabilities.

\subsection{Implications for Development and Testing of Code Generation Tools}

As discussed in Section \ref{sec:vuln-analysis}, we observed that there were certain vulnerability types for which Copilot security performance was diminished. As a result we suggest two approaches that CGT developers can consider as ways to improve the security performance of CGTs: targeted dataset curation and targeted testing. By targeted dataset curation, we mean specifically increasing the proportion of code samples that avoid a certain vulnerability type/CWE in the training data of a CGT in order to improve its performance with respect to that vulnerability. After training, targeted testing may also be used to test the CGT more frequently on the vulnerability types against which it performs poorly so that its strengths and weaknesses may be more accurately assessed.

\section{Threats to Validity}

\subsection{Construct Validity}


\textbf{Security Analysis.} Our recategorization (Section \ref{sec:recategorization}) relied on manual inspection by experts. While manual inspection has been used in other security evaluations of Copilot and language models \citep{pearce_asleep_2022} it makes it possible to miss other vulnerabilities that may be present. Our analysis resulted in a substantial number of category C samples ($\approx$42\%). We know little about the vulnerability level of these samples. Our analysis approach also does not allow us to determine whether other types of vulnerabilities (other than the original) may be present in the category A and B samples.

\textbf{Prompt Creation.} We re-create scenarios that led to the introduction of vulnerabilities by developers so that Copilot can suggest code that we can analyze. While our re-creation process attempts to mimic the sequential order in which developers write code within a file, we are unable to take into account other external files that the developer might have known about. As a result, Copilot may not have had access to the same amount/kind of information as the human programmer during its code generation. In spite of this, we see Copilot producing the fix in approximately 25\% of cases.

\subsection{Internal Validity}

\textbf{Training Data Replication.} The suggestions that Copilot makes during this study are frequently an exact match with code from the projects reported in the dataset. Copilot seems to occasionally replicate the vulnerable code or the fixed code. This observation forms the basis for our conclusion that Copilot's performance varies with different vulnerabilities. Another possible explanation for this is that the samples in question are included in Copilot’s training data.
However, given that GitHub reported Copilot's training data copying rate at approximately 1\% \citep{github_inc_github_2021}, this explanation does not completely explain our observations in this study where the replication rate would be greater than 50\%.
Also, considering Copilot’s lack of preference for either the vulnerable or vulnerability-fixing code (even if both are in its training dataset), we believe the findings of this study set the stage for further investigation into Copilot's memorization patterns. Such investigations may either have to find ways to overcome the lack of access to Copilot's training data or pivot to open-source models and tools. 

\subsection{External Validity}

\textbf{CWE Sample size.}
Our focus on Copilot's overall performance resulted in uneven counts of samples for each CWE that we encountered. To enable further vulnerability analysis, we only focused on CWEs that had at least three results across both categories A and B. Future work on vulnerability-specific performance may be better served with a more targeted sampling method that selects greater counts for each CWE. 

\textbf{Programming Languages.}
The dataset used for this evaluation only contained C and C++ code meaning Copilot's performance in this study may not generalize to significantly different programming languages. The findings are still valid for C/C++ which are still widely used.

\textbf{Other CGTS.}
This study focuses on Copilot's security performance on a dataset of real world vulnerabilities. Copilot's performance in this study cannot be generalized to all other CGTs because different CGTs will have different training datasets and different architectures. However, considering that Copilot is a fairly advanced and widely popular tool, we believe that critiques of and improvements to Copilot will probably apply to other CGTs as well. 

\textbf{Copilot Performance.}
Due to the diverse nature in which CGTs like Copilot can be prompted, combined with their non-deterministic nature, it is difficult to assume that Copilot's performance with respect to the different vulnerabilities will always be the same. In this study, Copilot was used in autopilot mode without any additional user intervention. It is possible, for example, that Copilot being used as an assistant may lead to different results. Still, our findings can serve as a guide for developers and researchers in pointing out situations in which Copilot's leash should be tightened or relaxed. 

\section{Related Work}\label{sec:related}

\subsection{\textbf{Evaluations of Language Models}}
As mentioned earlier, Copilot is the most evolved and refined descendant of a series of language models, including Codex \citep{chen_evaluating_2021} and GPT-3 \citep{brown_language_2020}. Researchers have evaluated and continue to evaluate language models such as these in order to measure and gain insights about their performance. 

Chen et al. \citep{chen_evaluating_2021} introduced and evaluated the Codex language model which subsequently became part of the foundation for GitHub Copilot. Codex is a descendant of GPT-3, fine-tuned on publicly available GitHub code. It was evaluated on the HumanEval dataset which tests correctness of programs generated from docstrings. In solving 28.8\% of the problems, Codex outperformed earlier models such as GPT-3 and GPT-J which managed to solve 0\% and 11.4\% respectively. In addition to finding that repeated sampling improves Codex’s problem solving ability, the authors also extensively discussed the potential effects of code generation tools. 

Li et al. \citep{li_competition-level_2022}addressed the poor performance of language models (such as Codex) on complex problems that require higher levels of problem solving by introducing the AlphaCode model. They evaluated AlphaCode on problems from programming competitions that require deeper reasoning and found that it achieved a top 54.3\% ranking on average. This increased performance was owed to a more extensive competitive programming dataset, a large transformer architecture, and expanded sampling (as suggested by Chen et al.). 

Xu et al. \citep{xu_systematic_2022} performed a comparative evaluation of various open source language models including Codex, GPT-J, GPT-Neo, GPT-NeoX, and CodeParrot. They compared and contrasted the various models in an attempt to fill in the knowledge gaps left by black-box, high-performing models such as Codex. In the process, they also presented a newly developed language model trained exclusively on programming languages - PolyCoder. The results of their evaluation showed that Codex outperforms the other models despite being relatively smaller, suggesting that model size is not the most important feature of a model. They also d that training on natural language text and code may benefit language models based on the better performance of GPT-Neo (which was trained on some natural language text) compared to PolyCoder (which was trained exclusively on programming languages). 

Ciniselli et al. \citep{ciniselli_what_2022} measured the extent to which deep learning-based CGTs clone code from the training set at inference time. As a result of a lack of access to the training datasets of effective code CGTs like Copilot, the authors trained their own T5 model and used it to perform their evaluation. They found that CGTs were likely to generate clones of training data when they made short predictions and less likely to do so for longer predictions. Their results indicate that Type-1 clones, which constitute exact matches with training code, occur about 10\% of the time for short predictions and Type-2 clones, which constitute copied code with changes to identifiers and types, occur about 80\% of the time for short predictions. 

Yan et al. \citep{yan_whygen_2022} performed a similar evaluation of deep learning-based CGTs with the goal of trying to explain the generated code by finding the most closely matched training example. They introduced WhyGen, a tool they implemented on the CodeGPT model \citep{lu_codexglue_2021}. WhyGen stores fingerprints of training examples during model training which are used at inference time to ``explain’’ why the model generated a given output. The explanation takes the form of querying the stored fingerprints to find the most relevant training example to the generated code. WhyGen was reported to be able to accurately detect imitations about 81\% of the time. 

\subsection{\textbf{Evaluations of Copilot}}

Nguyen and Nadi \citep{nguyen_empirical_2022} conducted an empirical study to evaluate the correctness and understandability of code generated by Copilot. Using 33 questions from LeetCode (an online programming platform), they created queries for Copilot across four programming languages. Copilot generated 132 solutions which were analyzed for correctness and understandability. The evaluation for correctness relied on LeetCode correctness tests while the evaluation for understandability made use of complexity metrics developed by SonarQube. They found that Copilot generally had low complexity across languages and, in terms of correctness, performed best in Java and worst in Javascript. 

Sobania et al. \citep{sobania_choose_2022} evaluated GitHub Copilot’s program synthesis abilities in relation to approaches taken in genetic programming. Copilot was evaluated on a standard program synthesis benchmark on which genetic programming approaches had previously been tested in the literature. They found that while both approaches performed similarly, Copilot synthesized code that was faster, more readable, and readier for practical use. In contrast, code synthesized by genetic approaches was ``often bloated and difficult to understand’’.  

Vaithilingam et al. \citep{vaithilingam_expectation_2022} performed a user study of Copilot in order to evaluate its usability. Through their observation of the 24 participants in the study, they identified user perceptions of and interaction patterns with Copilot. One of the main takeaways was that Copilot did not necessarily reduce the time required to complete a task, but it did frequently provide good starting points that directed users (programmers) towards a desired solution. 

Dakhel, Majdinasab et al. \citep{dakhel_github_2022} also performed an evaluation of Copilot with two different approaches. First, they examined Copilot’s ability to generate correct and efficient solutions for fundamental problems involving data structures, sorting algorithms, and graph algorithms. They then pivoted to an evaluation that compared Copilot solutions with that of human programmers. From the first evaluation, they concluded that Copilot could provide solutions for most fundamental algorithmic problems with the exception of some cases that yielded buggy results. The second comparative evaluation showed that human programmers generated a higher ratio of correct solutions relative to Copilot. 

Barke et al. \citep{barke_grounded_2022} conducted a grounded theory evaluation that took a closer look at the ways that programmers interact with Copilot.  Their evaluation consisted of observing 20 participants solving programming tasks across four languages with the assistance of Copilot. They found that there were primarily two ways in which participants interacted with Copilot: acceleration and exploration. In acceleration mode, participants already had an idea of what they want to do and used Copilot to accomplish their task faster. In exploration mode, participants used Copilot as a source of inspiration in order to find a path to a solution. 

Ziegler et al. \citep{ziegler_productivity_2022} compared the results of a Copilot user survey with data directly measured from Copilot usage. They asked Copilot users about its impact on their productivity and compared their perceptions to the directly measured data. They reported a strong correlation between the rate of acceptance of Copilot suggestions (directly measured) and developer perceptions of productivity (user survey).

Pearce et al. \citep{pearce_asleep_2022} performed an evaluation of Copilot with a focus on security. They investigated Copilot’s tendency to generate insecure code by curating a number of prompts (incomplete code blocks) whose naive completion could have introduced various vulnerabilities. They tasked Copilot with generating suggestions/completions for these prompts and analyzed the results using a combination of CodeQL and manual inspection of source code. They found that Copilot, on its own, generates vulnerable suggestions about 40\% of the time.

In our study, we perform a security centered evaluation of Copilot. Most of the previously discussed works surrounding evaluation of CGTs (with the exception of \citep{pearce_asleep_2022}) generally tend to focus on their usability and correctness. We believe that security evaluations of these tools at an early stage are crucial for preventing AI code assistants, which are likely to become popular amongst developers, from becoming large scale producers of insecure code. Our evaluation uses a dataset of vulnerabilities to investigate whether Copilot will always perform as badly as human developers in cases where the latter have previously introduced vulnerable code.

\section{Conclusion}\label{sec:conclusion}

Based on our experiments, we answer our research question negatively, concluding that Copilot is not as bad as human developers at introducing vulnerabilities in code. We also report on two other observations: (1) Copilot is less likely to generate vulnerable code corresponding to newer vulnerabilities, and (2) Copilot is more prone to generate certain types of vulnerabilities. Our observations in the distribution of CWEs across the categories indicates that Copilot performs better against vulnerabilities with relatively simple fixes. Although we concluded that Copilot is not as bad as humans at introducing vulnerabilities, our study also indicates that using Copilot to fix security bugs is risky, given that Copilot did introduce vulnerabilities in at least a third of the cases we studied.

Delving further into Copilot's behavior is hampered by the lack of access to its training data. Future work that either involves Copilot's training data or works with a more open language model can help in understanding the behaviour of CGTs incorporating such models. For example, the ability to query previous versions of the language model with the same query will facilitate longitudinal studies regarding how the CGT performs with respect to vulnerabilities of different ages. Similarly, access to the training data of the model can shed light on the extent to which the model memorizes training data.

A natural follow-up research question is whether the use of assistive tools like Copilot will result in \emph{less secure} code. Resolving this question will require a comparative user study where developers are asked to solve potentially risky programming problems with and without the assistance of CGTs so that the effect of CGTs can be estimated directly.

\section{Acknowledgements}

We acknowledge the fruitful discussions we had with the reviewers and the feedback received from them that helped improve this work, which was supported in part by the David R. Cheriton Chair in Software Systems and WHJIL.

\section{Data Availability Statement}

The dataset used in this study is made available by its curators \citep{fan_cc_2020} in the \emph{MSR\_20\_Code\_vulnerability\_CSV\_Dataset} repository: \url{https://github.com/ZeoVan/MSR_20_Code_vulnerability_CSV_Dataset}.

\bibliography{MSR_report}

\end{document}